# Identification and Visualization of Correlation Structures in Large-Scale Power Quality Data

Max Domagk
Jan Meyer
TUD Dresden University of Technology
Dresden, Germany
max.domagk@tu-dresden.de

Marco Lindner
TransnetBW
Stuttgart, Germany

*Abstract*— Large-scale power quality (PQ) measurement campaigns generate vast amounts of multivariate data, in which systematic dependencies are difficult to identify using conventional analysis techniques. This paper presents a methodology for the automated analysis and visualization of correlation structures in large PQ datasets. Building on an existing framework, the approach is adapted for shorter observation periods and enhanced with aggregation and distance-based visualization techniques. Daily Spearman correlation coefficients are averaged via Fisher's z-transformation and aggregated across phases, parameters, and sites. The resulting correlation structures are visualized using hierarchical clustering and multidimensional scaling to reveal consistent and recurring relationships. The methodology is demonstrated using data from 85 measurement sites within the German transmission system.

*Index Terms*-- power quality, correlation analysis, harmonics, data visualization, clustering, multidimensional scaling

## I. Introduction

Correlations are a key tool for analysing relationships and dependencies in power quality (PQ) measurement data. Of particular interest are correlations between different PQ parameters at an individual measurement site, as well as correlations of the same parameter across multiple sites. Automatically identified correlations can provide valuable insights into systematic dependencies within the power system, support the root-cause analysis of anomalous behaviour, and enable an assessment of how specific effects propagate throughout the network. In addition, correlation analysis can support the optimization of measurement campaigns, for example, by identifying redundant measurement locations that exhibit a high degree of correlation.

The methodology presented in this paper adapts the approach introduced in [1] for long-term measurements to shorter measurement periods. Furthermore, it introduces new visualization techniques for comprehensive correlation analyses. The objective is to identify and systematically evaluate robust relationships in large-scale measurement campaigns, which would otherwise be difficult to detect due to the high number of possible parameter and site combinations.

## II. Case Study: Dataset

The measurement sites are located within the German transmission system and distributed across three voltage levels: 38 sites at 110 kV, 21 at 220 kV, and 26 at 380 kV. In total, 85 sites are measured across 50 substations, covering both individual feeders (e.g., transmission lines) and transformer busbars. All measurements are conducted using IEC 61000-4-30 Class A compliant measurement instruments. The analysed voltage quality parameters include voltage RMS (Urms), short-term flicker (Upst), voltage unbalance (UNB), total harmonic distortion (Uthd), and voltage harmonics of odd orders 3 to 15; current parameters comprise current RMS (Irms), total harmonic current (Ithc), and current harmonics of odd orders 3 to 15. In total, 20 PQ parameters are considered, corresponding to 58 unique parameter-phase combinations per site.

## III. Correlation Analysis Methodology

In general, two types of correlation are distinguished: autocorrelation and cross-correlation. Autocorrelation describes the correlation within a single time series between observation at different time intervals, e.g., the daily evolution of the $5^{th}$ harmonic across consecutive days. It is commonly used to identify recurring temporal similarities. Cross-correlation quantifies the similarity between two different time series, such as comparisons of different PQ parameters at the same measurement site or of the same PQ parameter across different measurement sites.

### A. Correlation Coefficients

The Spearman rank correlation coefficient is a nonparametric measure for assessing the association between two time series. It is computed by applying the Pearson correlation coefficient to the ranked values of the data. For paired observations $(x_i, y_i)$ the coefficient is given by:

$$r = \frac{\sum_i (R(x_i) - \bar{R}_x)(R(y_i) - \bar{R}_y)}{\sqrt{\sum_i (R(x_i) - \bar{R}_x)^2} \sqrt{\sum_i (R(y_i) - \bar{R}_y)^2}} \qquad (1)$$

Unlike the Pearson coefficient, Spearman's coefficient is not restricted to linear relationships and more generally captures monotonic dependencies, i.e., variables that consistently increase or decrease together [1].

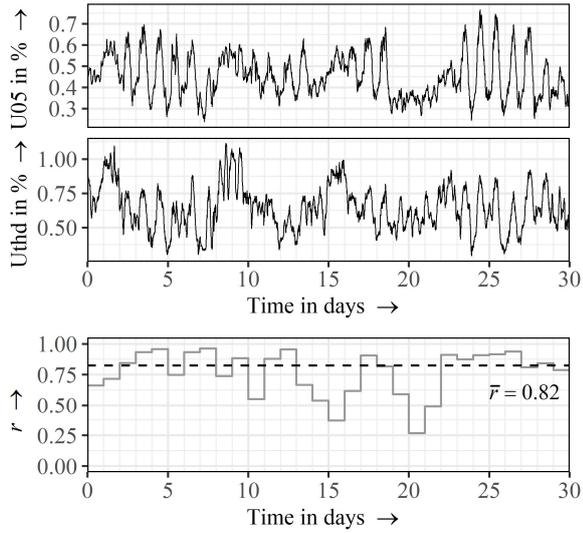

Figure 1. Time series of 10-minute average values for the 5th voltage harmonic and total harmonic voltage distortion (*top*); daily correlation coefficients between both time series (*bottom*) with the resulting mean correlation coefficient (*black dashed line*).

## B. Daily Correlation Coefficients

The calculation of daily correlation coefficients is illustrated using the 5th voltage harmonic (U05) and total harmonic distortion (Uthd) as an example. The corresponding time series are presented in Fig. 1 (top). Spearman correlation coefficients between these two PQ parameters were calculated daily over the entire measurement period, as shown in Fig. 1 (bottom; gray line). While the coefficients remain predominantly high, fluctuations are observed on specific days (e.g., days 16 and 21). This indicates that time-dependent factors can influence the correlation between PQ parameters. Consequently, correlations should not be assumed to be constant over the entire measurement duration, as this may lead to misinterpretation.

## C. Averaging Correlations

Mean correlation coefficients are calculated using the Fisher z-transformation to appropriately average daily coefficients and phase combinations. This two-stage process provides a robust estimate of the relationships between PQ parameters by compensating for the non-linearity of the correlation coefficient distribution [2].

A mean coefficient $\bar{r}$ is derived from $N$ daily coefficients as follows:

$$\bar{r} = \tanh\left(\frac{1}{N}\sum_{i=1}^{N} \operatorname{arctanh}(r_i)\right) \qquad (2)$$

Applying this to Fig. 1 yields $\bar{r} = 0.82$ (bottom; black dashed line), indicating a high correlation. Conversely, calculating a single correlation over the entire period of 30 days results in a lower value ($\bar{r} = 0.74$).

The mean coefficients are subsequently averaged across all phase combinations of two PQ parameters using the Fisher z-transformation. This step reduces the set of phase-specific correlations (e.g., L1-L1, L1-L2, ...) to a single representative

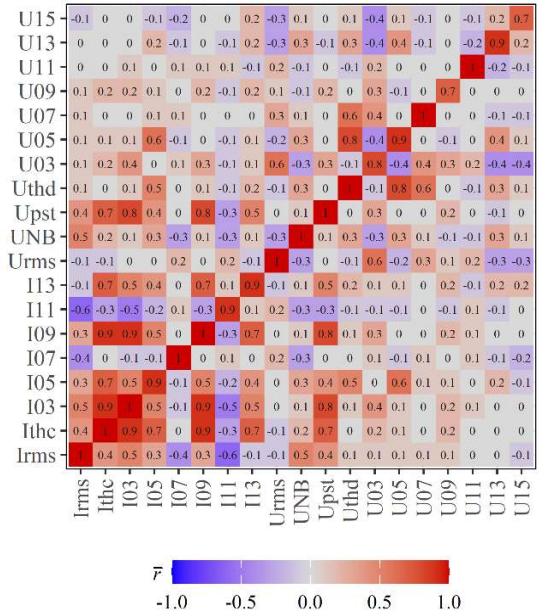

Figure 2. Correlation matrix of mean coefficients between power quality parameters at an EHV measurement site, considering all phase combinations; duration: 30 days

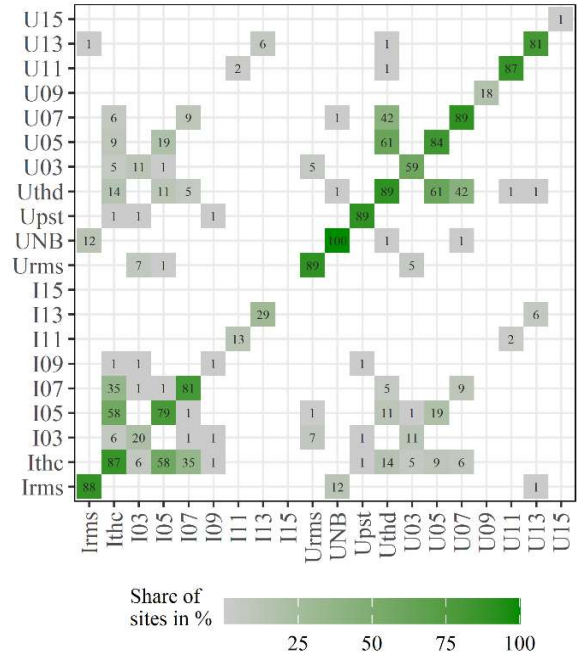

Figure 3. Share of measurement sites with significant correlation coefficients ($\bar{r} \geq 0.7$) between power quality parameters; total of 85 measurement sites from high-voltage (HV) and extra-high voltage (EHV) levels over a 30-day period

coefficient per parameter pair and site. As a result, the concise correlation matrix shown in Fig. 2 is obtained, which would otherwise be nine times larger. High values along the main diagonal confirm strong consistency between phases between the same PQ parameter. Notable results include strong correlations between current quality parameters (Ithc ~ I03 ~ I09) and flicker (Upst), while the total harmonic voltage distortion (Uthd) is primarily driven by the 5th harmonic (U05) at this measurement site.

## IV. Aggregation of Large Correlation Sets

Correlation matrices are exemplarily constructed between PQ parameters for multiple measurement sites. This results in a large number of correlation matrices for each measurement site, which complicates comprehensive analysis. In many cases, however, only significant correlations, particularly those with medium to high coefficients, are of interest. These are identified by applying a threshold $\tau_r$ to the mean correlation coefficients $\bar{r}_{ij}$ between two parameters $i$ and $j$, such that a correlation is considered significant if $\bar{r}_{ij} \geq \tau_r$. Depending on the analysis objective, the threshold can be applied to positive correlations, negative correlations ($\bar{r}_{ij} \leq -\tau_r$), or absolute values ($|\bar{r}_{ij}| \geq \tau_r$). Let $M$ denote the number of correlation matrices of dimensions $N \times N$. The aggregation matrix $\mathbf{A}$ is defined as

$$\mathbf{A} = \left[\frac{c_{ij}}{M}\right] \quad \text{with} \quad c_{ij} = \sum_{k=1}^{M} \mathbf{1}\left(\bar{r}_{ij}^{(k)} \geq \tau_r\right) \quad (3)$$

for $i,j = 1,2,\ldots,N$, where $\mathbf{1}(\cdot)$ denotes the indicator function. The resulting matrix provides a compact representation of how consistently significant correlations occur across all matrices and highlights robust correlation patterns.

Fig. 3 shows the relative occurrence of all significant correlations ($\bar{r} \geq 0.7$) between PQ parameters across 85 measurement sites. While some correlations observed at individual sites, such as Ithc ~ I09, occur only rarely, others, most notably Ihtd ~ I05, are consistently present. In contrast, correlations involving I09 and I15 do not exceed the threshold on average.

## V. Visualization of Correlation Structures

Large correlation matrices are difficult to interpret directly, particularly when analysing multiple PQ parameters across many measurement sites. Clustering and low-dimensional visualization are common tools to face this challenge, but they require first to be transformed into a distance representation.

### A. Distance Calculation

To obtain a distance matrix suitable for clustering and projection-based visualization, both correlation-based and aggregation-based representations are transformed using:

$$D_{ij} = \begin{cases} 1 - A_{ij}, & \text{for } i \neq j \\ 0, & \text{for } i = j \end{cases}. \quad (4)$$

$A_{ij}$ represents the aggregation value summarizing the consistency of significant correlations across multiple parameters or sites. When applied to a single correlation matrix, $A_{ij}$ corresponds directly to the correlation coefficient $\bar{r}_{ij}$, such that strongly correlated parameter pairs are mapped to small distances.

### B. Hierarchical Clustering Analysis

Hierarchical clustering is well-suited for analyzing large-scale correlation structures because it preserves pairwise relationships. In transient stability studies, correlation-based hierarchical clustering has been applied to identify generator coherency using rotor-angle correlations and synchronization coefficients [3].

In this study, aggregation-based distances are clustered using the average-linkage criterion. The resulting dendrogram (cf. Fig. 4) visualizes the share of significant correlations between voltage and current PQ parameters derived from the aggregation matrix in Fig. 3. Strong, frequent correlations appear as early merges at low linkage distances. The dendrogram reveals that significant correlations occur between total harmonic distortion and the 5th harmonic at more than half of the measurement sites, and with the 7th harmonic at over one third of the sites, for both voltage and current. Conversely, most other PQ parameters exhibit rare or no systematic correlation, with higher-order harmonics (e.g., 9th and 15th) remaining largely isolated.

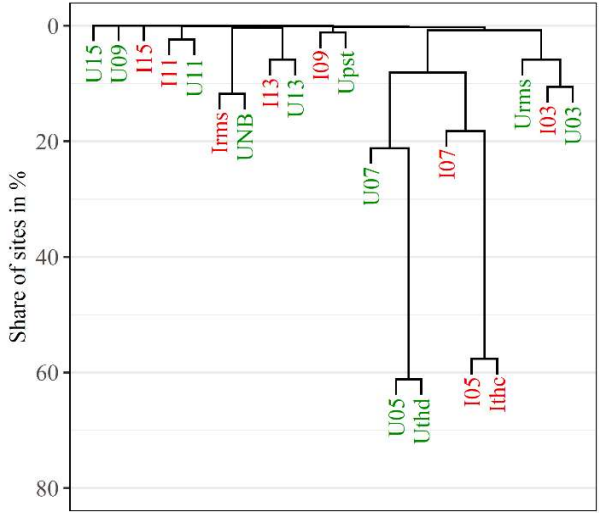

Figure 4. Visualization of the share of significant correlations between current (*red*) and voltage quality parameters (*green*) from Fig. 3 using the dendrogram of clustering based on the average-linkage method

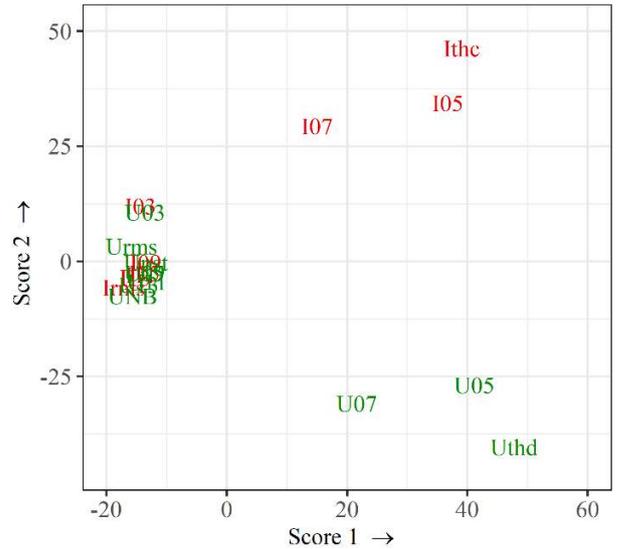

Figure 5. Visualization of the share of significant correlations between current (*red*) and voltage quality parameters (*green*) from Fig. 3 using a two-dimensional representation via multidimensional scaling (MDS)

### C. Dimensionality Reduction

Multidimensional scaling (MDS) embeds the distance matrix into a low-dimensional space while preserving pairwise distances as accurately as possible. Originally introduced in [4], MDS has also been applied in power systems to visualize electrical distances between buses instead of geographic coordinates [5].

The two-dimensional MDS representation in Fig. 5 illustrates the relative positions of PQ parameters based on their aggregation-based distance derived from the aggregation matrix in Fig. 3. Total harmonic distortion and the 5th and 7th harmonics, respectively, for current and voltage are located close to each other, reflecting their strong and frequent correlations. Parameters with weak or absent correlations are positioned at larger distances but remain close to each other, confirming the hierarchical clustering results from a geometric perspective.

For very large matrices, typically exceeding 10,000 variables, classical MDS becomes computationally expensive. In such cases, nonlinear dimensionality reduction techniques such as t-distributed Stochastic Neighbor Embedding (t-SNE) and Uniform Manifold Approximation and Projection (UMAP) provide scalable alternatives. Recent approaches, including sketch-and-scale strategies [6], further reduce computational complexity and enable visualization of large-scale correlation structures. For the matrix sizes considered in this study, classical MDS remains computationally feasible.

## VI. CASE STUDY: CORRELATIONS BETWEEN MEASUREMENT SITES

The methodology for identifying and visualizing correlation structures presented in Sections III to V has been demonstrated using correlations between PQ parameters at individual measurement sites. This case study extends the analysis by examining correlations between different measurement sites.

The correlation analysis between measurement sites is conducted for 20 parameters. Analogous to the parameter-based analysis, the share of significant correlations ($\bar{r} \geq 0.7$) is computed for all pairs of measurement sites. The resulting aggregation matrix for 85 sites is shown in Fig. 6. Due to the matrix size, direct interpretation is difficult. Therefore, a two-dimensional MDS representation (Fig. 7) and hierarchical clustering (Fig. 8) are used to visualize the correlation structure.

Based on these representations, three exemplary groups of measurement sites can be distinguished. Group 1 comprises sites that exhibit no or only very few correlations with other sites. These sites appear on the left side of the dendrogram and are located near the center of the two-dimensional representation. Group 2 includes measurement sites that show a higher mutual share of significant correlations, between 20 % and 60 % of the PQ parameters. These sites are predominantly located in the extra-high-voltage transmission level (220 kV and 380 kV) and appear relatively isolated from the remaining sites in both visualizations. Group 3 consists of measurement sites from the high-voltage level (110 kV), which also exhibit pronounced mutual correlations, affecting 40 % to 75 % of the PQ parameters, while showing weaker correlations with other groups. These sites are positioned centrally in the dendrogram and in the lower-left region of the MDS representation.

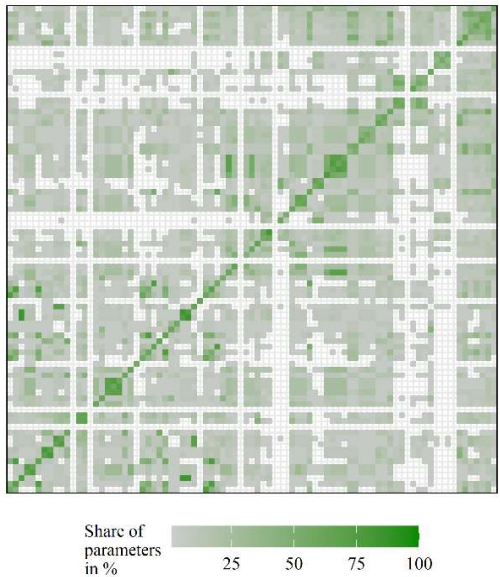

Figure 6. Share of parameters with significant correlation coefficients ($\bar{r} \geq 0.7$) across 85 measurement sites; 20 PQ parameters; 30-day period

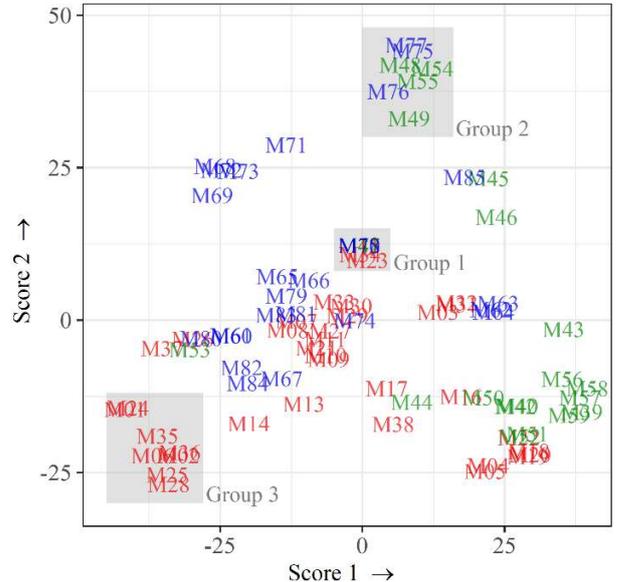

Figure 7. Visualization of the share of significant correlations across 85 measurement sites from Fig. 6 using a two-dimensional representation via multidimensional scaling (MDS); voltage levels: 110 kV (*red*), 220 kV (*green*) and 380 kV (*blue*)

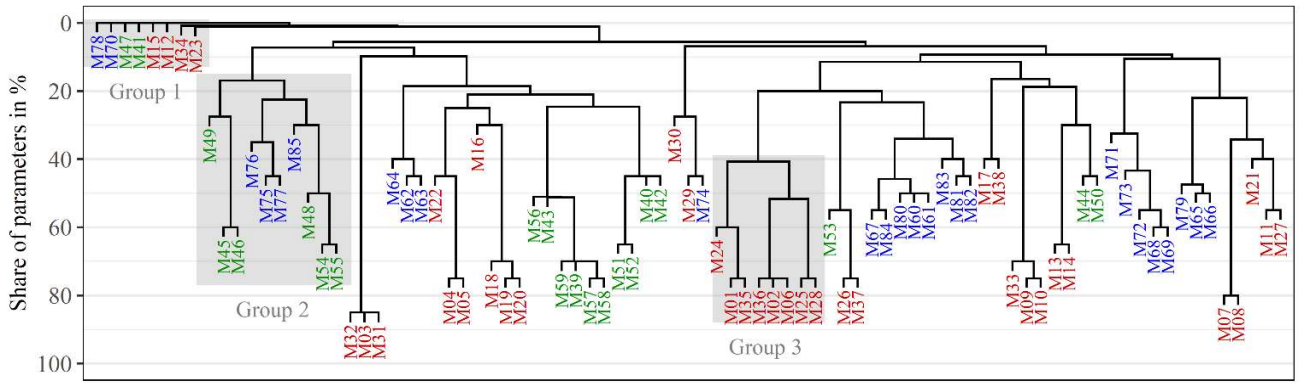

Figure 8. Visualization of the share of significant correlations across 85 measurement sites from Fig. 6 using the dendrogram of clustering based on the average-linkage method; voltage levels: 110 kV (*red*), 220 kV (*green*) and 380 kV (*blue*)

As indicated by the dendrogram, most measurement sites exhibit correlations for at least one or two PQ parameters, representing a 5-10 % share. Fig. 9 shows the share of significant correlations across all 85 measurement sites for each PQ parameter. The highest shares occur for voltage quality parameters U03 (47 %) and Upst (33 %), indicating that variations of these parameters tend to propagate over large parts of the network. The time series for the 3$^{rd}$ voltage harmonic across all sites (cf. Fig. 10) confirms these strong similarities in temporal characteristics.

## VII. Conclusions

This paper introduced a methodology for analyzing large-scale power quality correlation structures using robust averaging and aggregation techniques. By using hierarchical clustering and multidimensional scaling, complex dependencies become interpretable, revealing that low-order harmonics and flicker exhibit widespread spatial correlations, whereas higher-order harmonics remain largely localized. The approach enables systematic comparison across parameters and measurement sites and supports the identification of robust, recurring correlation patterns. The presented case studies demonstrate the applicability of the methodology correlation analyses in large measurement campaigns. Future research will focus on correlation-based segmentation to identify temporal changes, operational regimes, and periods of similar power quality behavior.

ACKNOWLEDGMENT

The authors would like to thank the Deutsche Forschungsgemeinschaft (DFG, German Research Foundation) for funding this project (project number 521923789).

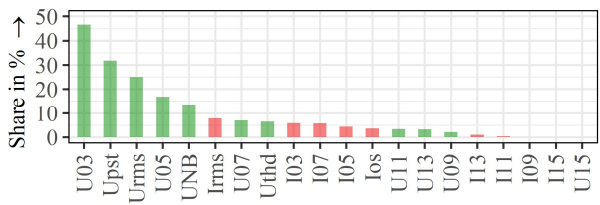

Figure 9. Share of significant correlations ($\bar{r} \geq 0.7$) among the $85^2$ site-to-site combinations for current (*red*) and voltage quality (*green*) parameters

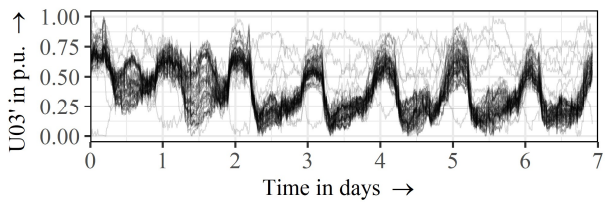

Figure 10. Example time series of 10-minute average values for the 3$^{rd}$ voltage harmonic across all 85 measurement sites; min-max normalized

References

[1] M. Domagk, J. Meyer, T. Wang, D. Feng, and W. Huang, 'Automatic Identification of Correlations in Large Amounts of Power Quality Data from Long-Term Measurement Campaigns', in *2021 26th International Conference and Exhibition on Electricity Distribution (CIRED)*, Online Conference,: Institution of Engineering and Technology, Nov. 2021, pp. 911–915. doi: 10.1049/icp.2021.1489.

[2] R. A. Fisher, 'Frequency Distribution of the Values of the Correlation Coefficient in Samples from an Indefinitely Large Population', *Biometrika*, vol. 10, no. 4, Art. no. 4, 1915, doi: 10.2307/2331838.

[3] F. Znidi, H. Davarikia, and H. Rathore, 'Power Systems Transient Stability Indices: Hierarchical Clustering Based Detection of Coherent Groups Of Generators', in *2024 IEEE Texas Power and Energy Conference (TPEC)*, Feb. 2024, pp. 1–6. doi: 10.1109/TPEC60005.2024.10472280.

[4] W. S. Torgerson, 'Multidimensional scaling: I. Theory and method', *Psychometrika*, vol. 17, no. 4, pp. 401–419, Dec. 1952, doi: 10.1007/BF02288916.

[5] F. Belmudes, D. Ernst, and L. Wehenkel, 'Pseudo-Geographical Representations of Power System Buses by Multidimensional Scaling', in *2009 15th International Conference on Intelligent System Applications to Power Systems*, Nov. 2009, pp. 1–6. doi: 10.1109/ISAP.2009.5352920.

[6] V. Wei, N. Ivkin, V. Braverman, and A. S. Szalay, 'Sketch and Scale Geo-distributed tSNE and UMAP', in *2020 IEEE International Conference on Big Data (Big Data)*, Dec. 2020, pp. 996–1003. doi: 10.1109/BigData50022.2020.9377843.